\documentclass[12pt]{iopart}
\usepackage{siunitx}
\DeclareSIUnit\pixel{px}
\usepackage{float}
\expandafter\let\csname equation*\endcsname\relax
\expandafter\let\csname endequation*\endcsname\relax
\usepackage[arrowdel]{physics}
\usepackage{upgreek}
\usepackage{orcidlink}

\expandafter\let\csname equation*\endcsname\relax
\expandafter\let\csname endequation*\endcsname\relax
\usepackage{amsmath}

\usepackage[symbol]{footmisc}

\begin{document}

\title[The role of flow field dynamics in enhancing VOC conversion in a Surface DBD system]{The role of flow field dynamics in enhancing volatile organic compound conversion in a surface dielectric barrier discharge system}








\author{Alexander Böddecker$^1$\orcidlink{0000-0002-2216-9515}, Maximilian Passmann$^2$\orcidlink{0000-0002-2403-6353}, Angie Natalia Torres Segura$^1$\orcidlink{0009-0008-7492-2389}, Arisa Bodnar$^1$\orcidlink{0009-0002-4506-9338}, Felix Awakowicz$^1$\orcidlink{0009-0000-2265-1144}, Timothy Oppotsch$^3$\orcidlink{0000-0003-1879-8690}, Martin Muhler$^3$\orcidlink{0000-0001-5343-6922}, Peter Awakowicz$^1$\orcidlink{0000-0002-8630-9900}, Andrew R. Gibson$^{4,5}$\orcidlink{0000-0002-1082-4359}, Ihor Korolov$^1$\orcidlink{0000-0003-2384-1243}, Thomas Mussenbrock$^1$\orcidlink{0000-0001-6445-4990}}

\address{$^1$ Chair of Applied Electrodynamics and Plasma Technology, Faculty of Electrical Engineering and Information Technology, Ruhr University Bochum,
Universitätsstraße 150, 44801 Bochum, Germany}
\address{$^2$ Chair of Hydraulic Fluid Machinery, Faculty of Mechanical Engineering, Ruhr University Bochum, Universitätsstraße 150,
44801 Bochum, Germany}
 \address{$^3$ Laboratory of Industrial Chemistry (LTC), Faculty of Chemistry and Biochemistry,
Ruhr University Bochum,
Universitätsstraße 150, 44801 Bochum, Germany}
\address{$^4$ Research Group for Biomedical Plasma Technology, Faculty of Electrical Engineering and Information Technology, Ruhr University Bochum, Universitätsstraße
150, 44801 Bochum, Germany}
\address{$^5$ York Plasma Institute, School of Physics, Engineering and Technology, University of York, Heslington, York YO10 5DD, United Kingdom}
\ead{alexander.boeddecker@rub.de}
\vspace{10pt}
\begin{indented}
\item[]\today


\end{indented}

\begin{abstract}

This study investigates the correlation between flow fields induced by a surface dielectric barrier discharge (SDBD) system and its application for the volatile organic compound (VOC) gas conversion process. As a benchmark molecule, the conversion of $n$-butane is monitored using flame ionization detectors, while the flow field is analysed using planar particle image velocimetry. Two individual setups are developed to facilitate both conversion measurement and investigation of induced fluid dynamics. Varying the gap distance between two SDBD electrode plates for three different $n$-butane mole fractions reveals local peaks in relative conversion around gap distances of \SI{16}{mm} to \SI{22}{mm}, indicating additional spatially dependent effects. The lowest $n$-butane mole fractions exhibit the highest relative conversion, while the highest $n$-butane mole fraction conversion yields the greatest number of converted molecules per unit time. Despite maintaining constant energy density, the relative conversion exhibits a gradual decrease with increasing distances. The results of the induced flow fields reveal distinct vortex structures at the top and bottom electrodes, which evolve in size and shape as the gap distances increase. These vortices exhibit gas velocity magnitudes approximately seven times higher than the applied external gas flow velocity. Vorticity and turbulent kinetic energy analyses provide insights into these structures' characteristics and their impact on gas mixing. A comparison of line profiles through the centre of the vortices shows peaks in the middle gap region for the same gap distances, correlating with the observed peaks in conversion. These findings demonstrate a correlation between induced flow dynamics and the gas conversion process, bridging plasma actuator studies with the domain of chemical plasma gas conversion. 



\end{abstract}

%
%
%
%
\ioptwocol

\section{Introduction}
\label{sec:intro}

Dielectric barrier discharges (DBDs) ignite as microdischarges under atmospheric pressure conditions and can be operated over a wide temperature range \cite{kogelschatzDielectricBarrierDischargesPrinciple1997a}. Kogelschatz \cite{kogelschatzPhysicsApplicationsDielectricbarrier2000} referred to this discharge type as the most important non-equilibrium discharge for atmospheric pressure. Under atmospheric pressure conditions, the resulting electron energies are well suited for dissociating and exciting gas molecules, which is promising for several applications. DBDs are commonly used for industrial ozone production \cite{ishidaAppearanceDCDielectric2011}, surface treatment, as well as pollution control \cite{kogelschatzDielectricBarrierDischargesPrinciple1997a, kogelschatzDielectricBarrierDischargesTheir2003a}. One important application of pollution control is the treatment of volatile organic compounds (VOCs) in the ambient air. Exposure to VOCs can increase the risk of several human diseases, such as cancer and birth defects in children and is, therefore, a relevant part of the hazardous global pollution emission \cite{heusinkveldNeurodegenerativeNeurologicalDisorders2016a,pereraEarlylifeExposurePolycyclic2014a, sarigiannisExposureMajorVolatile2011a, cosselmanEnvironmentalFactorsCardiovascular2015a}. Because of the need to preserve the human health, national and international bodies, such as the European Council, set up laws to control VOC emissions in the industry sector \cite{Directive200850}. While conventional thermal systems for the VOC decomposition rely on fossil fuels \cite{lewandowskiDesignThermalOxidation2017} and, additionally, are extremely energy intensive \cite{warahenaEnergyRecoveryEfficiency2009}, the DBD technology is a promising alternative. There, the generated reactive species in the active discharge region, namely electrons, ions and excited species, convert the pollutants at low gas temperatures efficiently. Furthermore, DBDs can be powered by electricity produced from renewable energy sources directly. 

Industrial processes require the treatment of high gas flow rates, making surface DBDs (SDBDs) a more suitable candidate compared to volume DBDs (VDBDs). SDBDs ignite on the surface of a dielectric material, which allows an easy scale-up and unlike VDBDs are not limited to small gas gaps, which would increase the flow resistance in such systems significantly \cite{boddeckerInteractionsFlowFields2023a}. Additionally, SDBDs have the potential to drastically influence the gas flow patterns,  similar to plasma actuator systems \cite{corkeDielectricBarrierDischarge2010}.   

Initially, the research on plasma actuators was motivated by their potential to improve the aerodynamics of planes or fluid machinery by active flow control. This topic has been studied for several decades \cite{corkeDielectricBarrierDischarge2010, jonathanPlasmaAerodynamicsCurrent2015}. During the discharge operation, the ionized gas exerts a body force on the surrounding neutral air, commonly known as the ionic wind \cite{robinsonHistoryElectricWind1962a}. This ionic wind actively manipulates aerodynamics, including the control of the boundary layer at low Reynolds numbers \cite{greenblattMechanismFlowSeparation2012, fengFlowControlAirfoil2015, satoMechanismsLaminarSeparatedflow2015}. \\
While the detailed processes are complex, the observed effects depend on several parameters, such as the voltage waveform \cite{qiPlasmaActuatorPerformance2017, borghiPlasmaAerodynamicActuator2015}, gas type \cite{jayaramanModelingDielectricBarrier2008, singhModelingPlasmaActuators2007}, surface effects \cite{erfaniPlasmaActuatorInfluence2012} and electrode geometry \cite{hoskinsonDifferencesDielectricBarrier2010, frizEffectsElectrodeSize2014}. Our SDBD source differs from conventional actuator designs by its large discharge area. Initially, it was constructed for chemical gas conversion purposes exclusively. It is well studied regarding its basic discharge parameters \cite{kogelheideCharacterisationVolumeSurface2020a, offerhausSpatiallyResolvedMeasurements2017b, nguyen-smithMsNsTwin2022a} and its VOC decomposition performance \cite{schuckeConversionVolatileOrganic2020a, schuckeOpticalAbsorptionSpectroscopy2022a, boddeckerScalableTwinSurface2022b, schuckeAnalysisReactionKinetics2022,petersCatalystenhancedPlasmaOxidation2021}. Böddecker \textit{et al.} \cite{boddeckerScalableTwinSurface2022b} observed a notably high conversion efficiency in their study, where they utilized a scaled-up SDBD system featuring multiple parallel-operated SDBD plates. They proposed that the presence of complex fluid structures could explain this result. In a recent study, Böddecker \textit{et al.} \cite{boddeckerInteractionsFlowFields2023a} have proven that this SDBD source generates fluid vortex structures analogous to the plasma actuator studies. The measurements were performed under quiescent air conditions.

In this paper, we aim to combine insights from actuator research and plasma-driven gas conversion. We utilize our previously used SDBD electrode design to investigate the plasma-induced fluid dynamics in the presence of an external gas flow, and compare these results with compatible VOC conversion results. We present results that show a correlation between the induced flow patterns and the conversion process through enhanced gas mixing.

\section{Experiment and Diagnostics}
\label{sec:exp}
\subsection{Surface dielectric barrier discharge generation}

\begin{figure}[hbt]
    \centering
    \includegraphics[width= \linewidth]{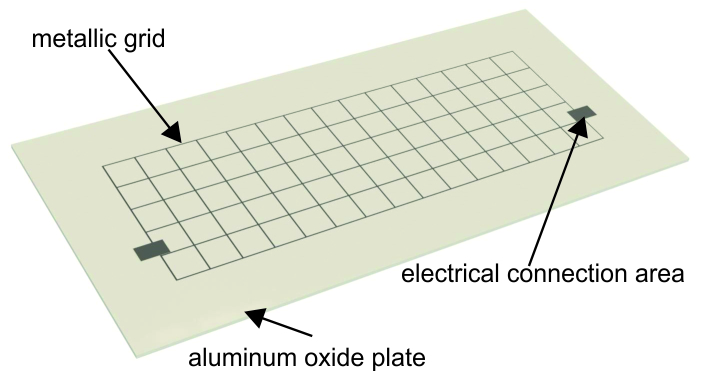}
	\caption{Drawing of the used SDBD electrode.}
	\label{fig:electrode}
\end{figure}

Figure \ref{fig:electrode} displays a drawing of our used SDBD model for this work - fabricated by Alumina Systems GmbH, Germany. The dielectric plate made of aluminium oxide ($\mathrm{\upalpha}$-Al$_2$O$_3$) serves to prevent the discharge from transitioning to a thermal arc and has a size of \SI{190}{mm} $\times$ \SI{88}{mm} $\times$ \SI{0.635}{mm}. The anode and cathode are silk-screen printed onto both sides of the plate in the form of a quadratic grid with a constant grid spacing of \SI{10}{mm}. The opposing grid on the back side of the plate is congruent to the visible one on the top side. The grid lines are \SI{0.45}{mm} wide and consist of molybdenum and manganese silicate (80/20) and are chemically nickelled. The total grid size is \SI{150.45}{mm} $\times$ \SI{50.45}{mm} and contains 75 cells in total. To prevent surface discharges around the edges, the grid maintains a distance of \SI{14.8}{mm} on the long side and \SI{18.8}{mm} on the short side from the plate's edge. Electrical connection areas on both sides of the plates facilitate basic contacting through soldering tin or spring plate contacts.  

In our experiment, one side of the grid is permanently connected to the ground potential, and the other side is powered with a high-voltage (HV) signal. The setup enables the simultaneous operation of multiple parallel electrode plates, facilitating the study of their spatially dependent induced flow field interactions. This multi-electrode configuration also serves as a model for upscaled systems designed for VOC remediation of high gas flow rates. The two opposing SDBD plates are arranged with an asymmetrical polarity configuration, resulting in opposing ground potential sides for both the conversion and PIV experiments. The voltage signal is produced by a G2000 plasma generator (Redline Technologies, Germany). This power supply generates unipolar square pulses with voltage amplitudes up to \SI{300}{V} and repetition frequencies $f_{\mathrm{rep}}$ up to \SI{500}{kHz}. An external transformer steps up this voltage, exciting the natural resonance circuit formed by the transformer's inductance and the SDBD electrode's capacitance. The resulting voltage waveform is a damped sine wave with a resonance frequency in the range of several kHz. Its peak-to-peak amplitude of the first two halfwaves is set constant to \SI{11}{kV} for all measurements performed in this work. This leads to the possibility of igniting several discharge cycles per pulse period. For a more detailed description of the electrical setup and its use, we refer to the works of Schücke \textit{et al.} \cite{schuckeConversionVolatileOrganic2020a, schuckeAnalysisReactionKinetics2022}, Nguyen-Smith \textit{et al.}  \cite{nguyen-smithMsNsTwin2022a} and Böddecker \textit{et al.} \cite{boddeckerScalableTwinSurface2022b}.

For the measurement campaigns in this study, it is important to control the electrical power given by
\begin{gather}
    P = f_{\mathrm{rep}}\int_0^{T_{\mathrm{p}}} U(t)  \left[ I(t) - C  \dv{U(t)}{t} \right]  \mathrm{d}t.
    \label{Eq:Power}
\end{gather}
for achieving comparable results.
Here, $f_{\mathrm{rep}}$ is the repetition frequency, $T_{\mathrm{p}}$ denotes the duration of the pulse, $U(t)$ stands for the time-resolved voltage, $I(t)$ is the time-resolved electrical current, $C$ denotes the capacitance of the system and $t$ is the time. The term $C\mathrm{d}U/\mathrm{d}t$  is the displacement current that does not have a net power contribution and is subtracted accordingly. As established in previous studies, this term is safely disregarded, facilitating faster power calculations. This allows a real-time power calculation and control by adjusting the repetition frequency while keeping voltage amplitudes constant.

The electrical current is measured using a current probe (Model 6585, Pearson Electronics, USA), while the voltage is measured with a high-voltage probe (P6015A, Tektronix GmbH, Germany). Both signals are displayed and recorded on oscilloscopes, specifically a Waverunner 8254 (Teledyne LeCroy, USA) for the conversion experiment and an HDO6104A (Teledyne LeCroy, USA) for the PIV setup.

\subsection{Conversion experiment}

\begin{figure*}[hbt]
    \centering
    \includegraphics[width=\textwidth]{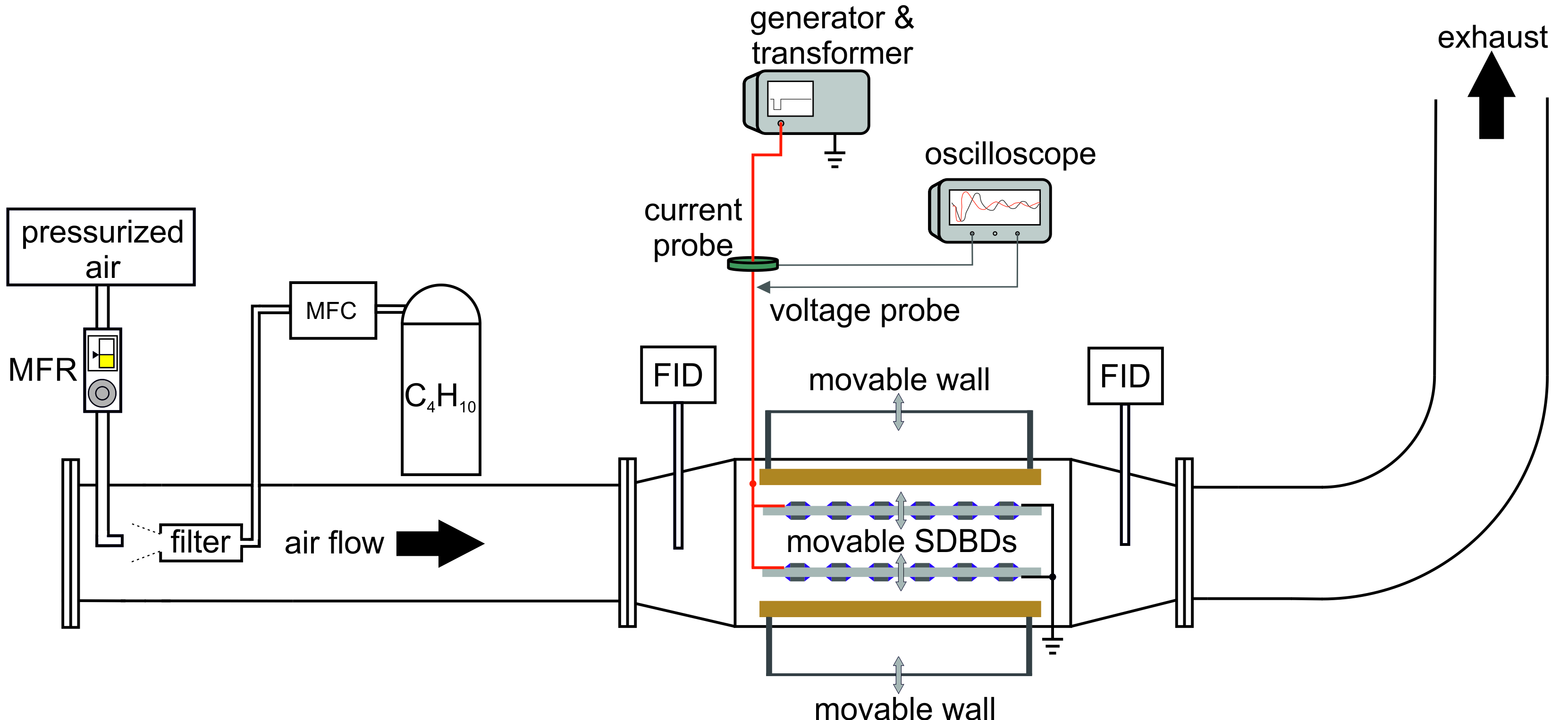}
	\caption{ Schematic drawing of the experimental setup for measuring the $n$-butane conversion adapted from \cite{boddeckerScalableTwinSurface2022b}.  }
	\label{fig:conversion_setup}
\end{figure*}

\begin{figure}[hbt]
    \centering
    \includegraphics[width = \linewidth]{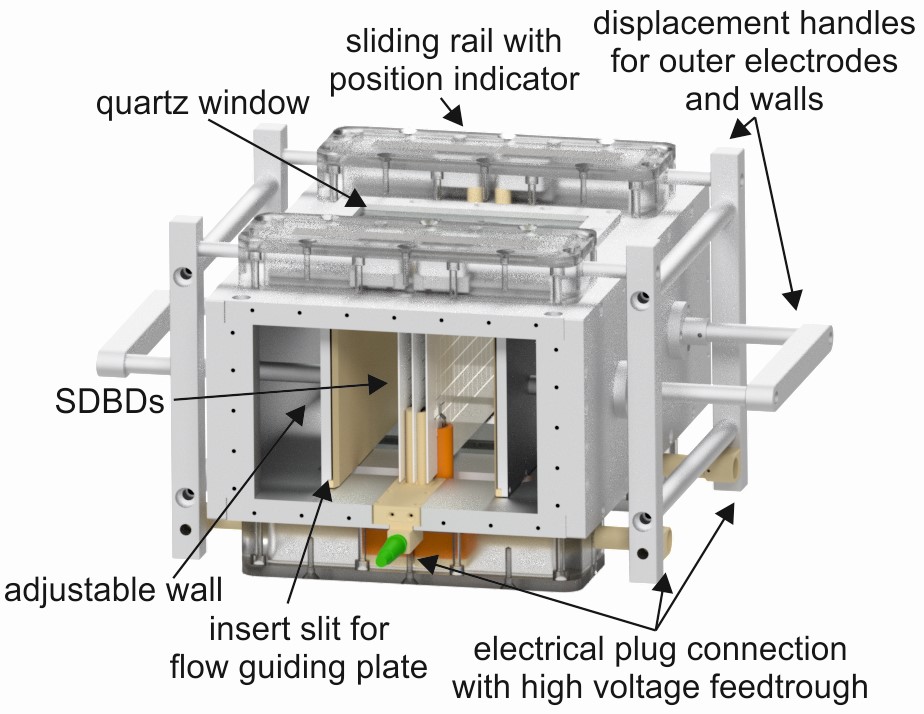}
	\caption{Frontal depiction of the 3D model of the used SDBD chamber for the conversion experiments. } 
	\label{fig:conversion_chamber}
\end{figure}

For measuring the conversion of $n$-butane, a similar setup as described in Böddecker \textit{et al.} \cite{boddeckerScalableTwinSurface2022b} is used. A schematic drawing is displayed in figure \ref{fig:conversion_setup}. The pipe system (Pfeiffer Vacuum Technology AG, Germany) for the primary gas flow has an inner diameter of \SI{70}{mm}. The air flow is controlled with a mass flow regulator (MASS-VIEW MW-308, Bronkhorst Deutschland Nord GmbH, Germany) and uses oil-filtered pressurized air. The presented system is able to deliver \SI{0}{slm}-\SI{100}{slm}, which is defined under normal conditions by the company at \SI{0}{\degreeCelsius} and \SI{1013.25}{mbar}. For all measurements presented in this study, the flow velocity is set constant to \SI{0.1}{ms^{-1}}. The resulting Reynolds number is $\text{\textit{Re}}=$ 466. The flow can be therefore assumed to be a laminar pipe flow.  For all access lines and measurement lines, \SI{6}{mm} (outer diameter) stainless steel tubing and connectors (Hy-Lok D GmbH, Germany) are used. The end of the main gas stream piping is connected to the building exhaust to safely extract all residual chemicals out of the laboratory. $n$-Butane, which is our benchmark molecule for the VOC conversion in this study, is added to the main gas stream at the beginning of the main pipe system. It is injected via a mass flow controller (LOW-$\Delta$P-FLOW, Bronkhorst Deutschland Nord GmbH, Germany) through a particle filter (\SI{0.5}{\micro m} pore size, Swagelok Company, USA) for improved radial mole fraction distribution in the inner pipe. The $n$-butane gas is supplied by a gas cylinder (N25, AIR LIQUIDE Deutschland GmbH, Germany). Relative conversion of injected $n$-butane is measured using two flame ionization detectors (FID) (SmartFID ST, ErsaTec GmbH, Germany). The FIDs sample \SI{1.4}{slm} gas from the centre of the main pipe, combusting it in an H$_2$ flame between electrodes. The resulting ions induce an electrical current, proportional to the combusted VOC amount, measured via calibration with a known mole fraction calibration gas (CRYSTAL, AIR LIQUIDE Deutschland GmbH, Germany). All measurement and process data is stored with a frequency of \SI{10}{Hz} using a self programmed control program with LabVIEW (LabVIEW 2019, National Instruments Cooperation, USA). A time period of \SI{90}{s} is required until the process reaches a steady state. All parameters are recorded over a duration of \SI{120}{s}, with the conversion and energy density values averaged over the final \SI{30}{s} of the recording period, as in previous studies \cite{boddeckerScalableTwinSurface2022b, schuckeConversionVolatileOrganic2020a}.

Relative $n$-butane conversion $X_\mathrm{rel}$ is determined by comparing the initial $n$-butane mole fraction $y_\mathrm{in}$ with post-discharge treatment mole fraction $y_\mathrm{out}$ by 
\begin{gather}
    X_\mathrm{rel} = \frac{y_\mathrm{in}-y_\mathrm{out}}{y_\mathrm{in}} \cdot 100\,\%.   
\end{gather}
In this study, the initial $n$-butane mole fraction is varied and set to \SI{100}{ppm}, \SI{500}{ppm} and \SI{1000}{ppm} for experimental investigation. All conversion measurements are reproduced three times in total to achieve information about the statistical error, which is indicated by the standard deviation in the error bars in the results section.

The distinct feature of the used discharge chamber lies in its movable SDBD plates, as well as the movement of the artificial walls, which are independent of each other. A more detailed drawing of this chamber can be seen in figure \ref{fig:conversion_chamber}. While the chamber is designed to accommodate three SDBD electrodes simultaneously, the middle electrode is removed to increase the possible distance between the electrodes further. The gap distance $d$ between the SDBD electrodes and that between the adjustable wall and the adjacent electrode are always kept the same. The total width of the chamber is consequently $3 d$. With the height of this chamber of \SI{93}{mm} the cross-section is easily estimated, which is important for translating the volumetric flow rate $\dot{V}$ into the average flow velocity. In this study, the energy density $\rho_{\mathrm{E}}$ is kept constant for all distance variations. It is calculated as
\begin{gather}
    \rho_{\mathrm{E}} = \frac{P(f_{\mathrm{rep}})}{\dot{V}} = \frac{E}{V}.
\end{gather}
Due to the cancellation of time in the fraction, $\rho_{\mathrm{E}}$ is also equivalent to the energy $E$ per unit volume $V$. Since wider gap distances require higher volumetric flows in order to preserve a gas velocity of \SI{0.1}{ms^{-1}}, the repetition frequency is adjusted to keep $\rho_{\mathrm{E}}$ constant. By preserving this constant, any change in the relative conversion should be exclusively attributed to changes in fluid dynamics. We assume that under constant power, a constant number of reactive species is generated. However, due to the variation in volumetric flow with gap distance, their density would vary, potentially resulting in a modified relative conversion. For comparative purposes with thermal reactors, $\rho_{\mathrm{E}}$ is a suitable quantity, as it is related to process temperature in those reactors.

Guiding plates, which are inserted in the adapter part, avoid gas flow in the space between the adjustable walls of the chamber and the fixed outer wall. Quartz windows at the top and bottom sides allow optical access to the chamber. The positions of the electrode plates can be read at the position indicators on the top side of the chamber. Using a caliper externally applied to the chamber windows for measuring the wall positions resulted in increased uncertainty, raising the total gap distance uncertainty to \SI{0.5}{mm}.

\subsection{Particle image velocimetry setup}

\begin{figure*}[hbt]
    \centering
    \includegraphics[width=\textwidth]{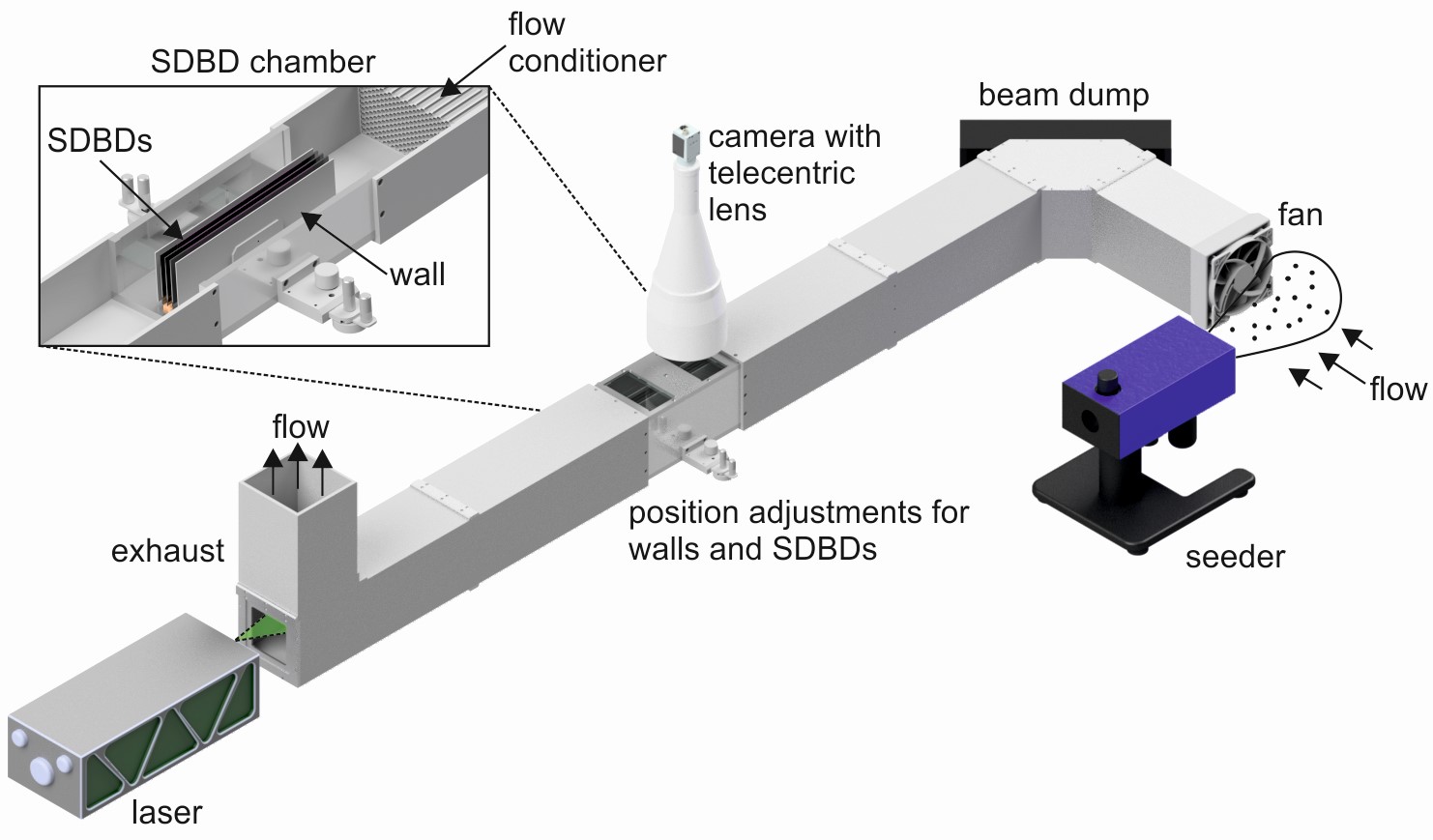}
	\caption{Image of the 3D model of the PIV experiment. A zoomed half-section view shows the completely equipped interior SDBD chamber as well as the used flow conditioner. }
	\label{fig:PIV_setup}
\end{figure*}


To correlate the conversion process with the underlying fluid dynamics, a separate setup is constructed, which has improved optical access for applying the fluid diagnostics.  
The optical setup is depicted in figure \ref{fig:PIV_setup}. A zoomed half-section view of the central element shows the interior of the SDBD chamber as well as the used flow conditioner for maintaining a laminar inflow. The SDBD chamber is a simplified model of the chamber used for conversion experiments, because its general requirements are lower. In figure \ref{fig:PIV_setup} the fully accommodated chamber with three SDBD electrodes is shown, but the middle one is removed for the results presented in this study. The rectangular pipe shape is suiting the cross-section of the reactor chamber, avoiding adapter parts that would increase turbulence. 

The gas flow speed is set by a PWM-fan and generates average air flow velocities between \SI{0.1}{ms^{-1}} to \SI{1}{ms^{-1}}.  
As an optical diagnostic for flow visualization, planar (2D-2C) particle image velocimetry (PIV) is used. The diagnostic setup (Multi-Parameter PIV / BOS / LIF System, LaVision GmbH, Germany) is commercially available. As seeding particles, DEHS (di-ethyl-hexyl-sebacate) droplets, in the order of \SI{1}{\micro m} in diameter, are introduced into the flow through an aerosol generator (LaVision GmbH, Germany). As a light source, a Nd:YAG double pulse laser (Evergreen 100, Quantel laser by LUMIBIRD, France) with pulse widths of 2 $\times$ \SI{8}{ns}, pulse energies in the range of 2 $\times$ \SI{100}{mJ}, a wavelength of \SI{532}{nm} and a maximum repetition rate of \SI{25}{Hz} is applied. The emitted light is formed into a horizontal light sheet by use of light sheet optics and travels through the pipe system until it is captured by a beam dump (LaVision GmbH, Germany) to avoid strong reflections and maintain laser safety. The particle scattering images are recorded with a laser-synchronized camera (Imager \textit{CX}-12 LaVision GmbH, Germany) with a resolution of 4080$\times$2984 pixels and dual frame technology. As the camera lens a high-resolution telecentric lens (TC1MHR080-C, Opto Engineering Deutschland GmbH, Germany) with a magnification of 0.134 and a working distance of \SI{226.8}{mm} is utilized for avoiding optical distortion in the measurement section. Due to the three-dimensional flow field profile, the particle losses at the imaging level of the exact position of a horizontal grid line are too significant to achieve quantitative PIV results. Therefore, to minimize the losses, the laser sheet is positioned vertically at a position, where it travels through the middle of two horizontal grid lines of the electrodes.  

The PIV analysis is performed with the DaVis software (Version 10, LaVision GmbH, Germany). The spatial calibration could be facilitated by using the grid constant of the electrodes, which is visible in the raw images and has a high precision due to the tight manufacturing tolerances. After this process step, a background image without seeding particles is recorded, which is subtracted by the software for all following PIV recordings to enhance the particle image contrast. Multi-pass cross-correlation with 2 initial passes of 128 $\times$ 128 pixels and a \SI{75}{\%} overlap and 2 final passes of 48 $\times$ 48 pixels with an overlap of \SI{50}{\%} was used in processing the PIV recordings. The final window size ensured at least 10 particle images per interrogation window [38]. The selected time-delay between the first and second particle images was selected as $\Delta t = \SI{400}{\micro s}$. The resulting particle displacements fulfilled the "one-quarter-rule" during the turned-on case, which means that the interrogation window size is four times larger than the mean particle displacement between two images. This minimizes in-plane losses, which would otherwise negatively impact the correlation performance \cite{scharnowskiParticleImageVelocimetry2020}. 

For comparing the flow fields at different operating parameters, we calculate its vorticity $\vec{\omega}$ in 2D according to \cite{kunduFluidMechanics2012} as
\begin{gather}
        \vec{\omega} = \vec{\nabla}\times{\vec{u}} =  \left( \frac{\partial u_\text{y}}{\partial x} - \frac{\partial u_\text{x}}{\partial y} \right) \vec{e}_{\mathrm{z}}  
        \label{eq:vorticity}
\end{gather}

In the context of our study, $\vec{u}=u_\text{x}\vec e_\text{x} +u_\text{y}\vec e_\text{y}$ represents the velocity vector, where $x$ and $y$ are the horizontal and vertical Cartesian coordinates, respectively. Additionally, $\vec{e}_{\mathrm{z}}$ denotes the unit vector perpendicular to the $xy$-plane. In two dimensions, this term simplifies to a scalar value representing the out-of-plane rotation, which is disregarded in this context. 

The vorticity directly identifies and quantifies vortices, providing insights into the strength of rotational motion and aiding in the prediction of boundary layer development \cite{esmailiTemporalDevelopmentTurbulent1992, pintado-patinoRoleInfiltrationExfiltration2015}, instabilities \cite{magnaudetWakeInstabilityFixed2007, lorenzaniFluidInstabilitiesPrecessing2001}, and recirculation zones \cite{al-abdeliRecirculationFlowfieldRegimes2003, oliveiraDivergentStreamlinesFree2012, kunduFluidMechanics2012}. Because the vorticity contains negative and positive values, indicating the rotational orientation of the vortices, we decided to use the absolute vorticity ${|\vec\omega|}$ in certain results to illustrate the total vortices intensity of the flow field, independent of its rotational direction. 

Another important quantity used in this study, commonly used for describing turbulent flow fields, is the turbulent kinetic energy (TKE). It quantifies the energy associated with turbulent motion in a flow.  Higher local TKE values indicate higher turbulence intensity zones, and higher turbulence lead to enhanced gas mixing \cite{oerteljrPrandtlFuehrerDurch}. In our 2D case, it is calculated according to  \cite{stullIntroductionBoundaryLayer1988} by
\begin{gather}
    \mathrm{TKE}  =\frac{1}{2} m  \left(  \overline{\delta u_\text{x}^2} + \overline{\delta u_\text{y}^2} \right)    
\end{gather}

Here, $m$ is the mass of the fluid, $\delta u_\text{x}^2$ and $\delta u_\text{y}^2$ denote the squared fluctuating velocity components in $x$- and $y$ direction given by

\begin{align}
     \overline{\delta u_\text{x}^2} &= \frac{1}{T} \int_0^T \delta u_\text{x}^2 \mathrm{d}t=\frac{1}{T} \int_0^T(u_\text{x}(t)-\overline{u_\text{x}})^2 \mathrm{d}t, \\
        \overline{\delta u_\text{y}^2} &= \frac{1}{T} \int_0^T \delta u_\text{y}^2 \mathrm{d}t= \frac{1}{T} \int_0^T(u_\text{y}(t)-\overline{u_\text{y}})^2 \mathrm{d}t.
\end{align}
 $t$ is the time and $T$ the duration of the recording in steady state conditions \cite{stullIntroductionBoundaryLayer1988}. The fluctuating velocity components are calculated as the difference between its instantaneous value ($u_\text{x}(t)$, $u_\text{y}(t)$) and its time-averaged velocity ($\overline{u_\text{x}}$, $\overline{u_\text{x}}$). In the absence of discharge operation $\delta u_\text{x}$ and $\delta u_\text{y}$ are zero. When calculating a kinetic energetic quantity, such as the TKE, the mass of the fluid must be taken into account.  For simplicity reasons, including this study, it is often neglected to show relative changes. Therefore, all TKE plots presented here are shown without consideration for mass.

In plasma actuator studies, an enhanced gas mixing \cite{depurumohanVortexenhancedMixingActive2015, waitzEnhancedMixingStreamwise1997} for stream wise vortices was already shown \cite{satoNumericalInvestigationStreamwise2022}. In this work, our aim is to use the vorticity- and the TKE profiles to correlate their behaviour also with gas mixing, which is supposed to be an important additional effect for the conversion process.


\section{Results and Discussion}
\label{sec:results}

\subsection{Conversion of $n$-butane}
\label{sec:conversion_results}

\begin{figure}[hbt]
    \centering
    \includegraphics[width=\linewidth]{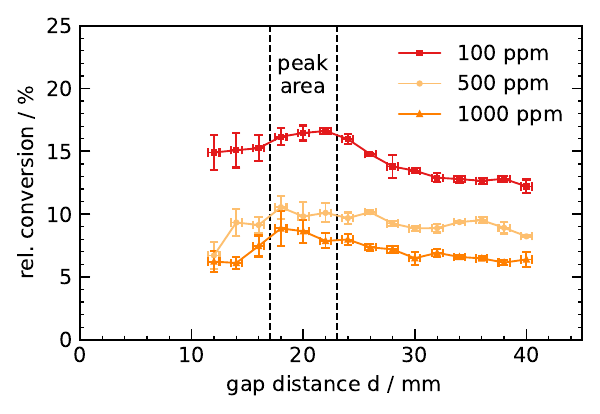}
	\caption{Measured relative conversion at different $n$-butane mole fractions as a function of the gap distance at a constant energy density of \SI{100}{JL^{-1}}. }
	\label{fig:conversion_comp}
\end{figure}

For the $n$-butane conversion measurements, a constant energy density of \SI{100}{JL^{-1}} is maintained for all conditions used in this work. Additionally, adjustments in the gap distance and consequently the chamber's cross-section require an adaption of the volumetric flow to keep the average gas flow velocity constant to \SI{0.1}{ms^{-1}}. Therefore, for each specific gap distance, a unique repetition frequency is used to keep the energy density constant. Maintaining a constant peak-to-peak voltage of \SI{11}{kV} ensures that the discharge dynamics are consistent across different gap distances. During the measurement campaign, the uncertainty of the energy density is approximately \SI{1.3}{JL^{-1}}, representing a relative error of \SI{1.3}{\%}.

Figure \ref{fig:conversion_comp} shows the relative conversion as a function of the gap distance for three applied $n$-butane mole fractions. The highest relative conversion of \SI{17.06}{\%} is achieved for a mole fraction of \SI{100}{ppm} at a gap distance of \SI{20}{mm} with consideration of the error bars. The relative conversion profiles are decreasing with higher $n$-butane mole fractions, confirming previous research findings \cite{boddeckerScalableTwinSurface2022b, schuckeConversionVolatileOrganic2020a}. The absolute conversion, defined as the number of converted molecules per time, would be higher the other way around. For example, the highest relative conversion for \SI{500}{ppm} is \SI{11.48}{\%} at \SI{18}{mm}, which is roughly \SI{33}{\%} smaller than for the \SI{100}{ppm} case, but the density of $n$-butane molecules is 5 times larger, which results in a higher absolute conversion. 
Intuitively, one would anticipate a trend where relative conversion gradually decreases with increasing distances, as the ratio between discharge volume and gap volume diminishes. As the power is also higher at larger gap distances, resulting in the production of more reactive species, this trend should be counterbalanced. However, additional factors, such as temperature changes and complex gas chemistry and flow field patterns, limit this simplistic model, leading to a gradual decrease in the conversion. 
Despite this, a small local maximum is found for all mole fractions between \SI{16}{mm} and \SI{22}{mm} in the relative conversion. Due to the fact that the energy density is kept constant, but the spatial conditions changed, a change in the fluid dynamics is expected to be the reason for the relative conversion maximum that is observed. This question is investigated in the following subsection.

\subsection{Induced fluid dynamics}

\begin{figure}[hbt]
    \centering
    \includegraphics[width=\linewidth]{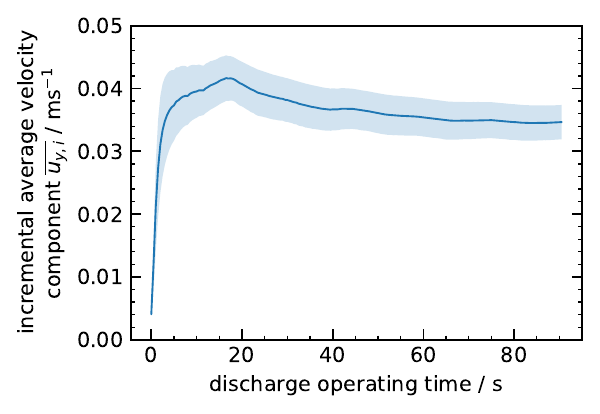}
	\caption{Convergence of the velocity component $u_\text{y}$ using the incremental average during the operating time. }
	\label{fig:convergence}
\end{figure}

In order to investigate the temporal convergence of the induced fluid dynamics, we conducted an initial test, with a recording time of \SI{90}{s}. To verify whether the flow field reached steady-state conditions and to ensure correct application of time-averaging, we use the incremental average formula for the spatially averaged vertical velocity component $\overline{u_{\text{y,i}}}$ per flow field number $i$. The flow field number refers to the number of recorded double images used for each PIV cross-correlation. The vertical velocity component is well-suited for this analysis, because it is close to \SI{0}{ms^{-1}} ($u_{\mathrm{y,off}}= \SI{4.7(1.7)}{mm \, s^{-1}}$) in the discharge off-state and only significantly present due to the flow induced by the discharges. The formula for the incremental average is given by
\begin{gather}
    \overline{u_{\text{y, i+1}}}  = \frac{ i\, \overline{u_{\text{y,i}}} + u_{\text{y,i+1}}}{i+1}.
\end{gather}
The result is depicted in figure \ref{fig:convergence}, illustrating typical convergence behaviour. The shaded region surrounding the line represents the incremental standard deviation, calculated analogously. A strong rise is observed during the first \SI{5}{s} of discharge operating time, followed by a peak at \SI{17}{s} where $\overline{u_{\text{y,i}}}$ begins to decay, stabilizing at a constant value of around \SI{0.034}{ms^{-1}}. Since the convergence reaches a steady state at \SI{80}{s},  the flow fields are recorded from \SI{80}{s} to \SI{90}{s}, with a repetition frequency of almost \SI{20}{Hz}, resulting in 189 images captured within \SI{10}{s} recording time. These flow fields are then time-averaged and analysed.

\begin{figure}[hbt]
    \centering
    \includegraphics[width=\linewidth]{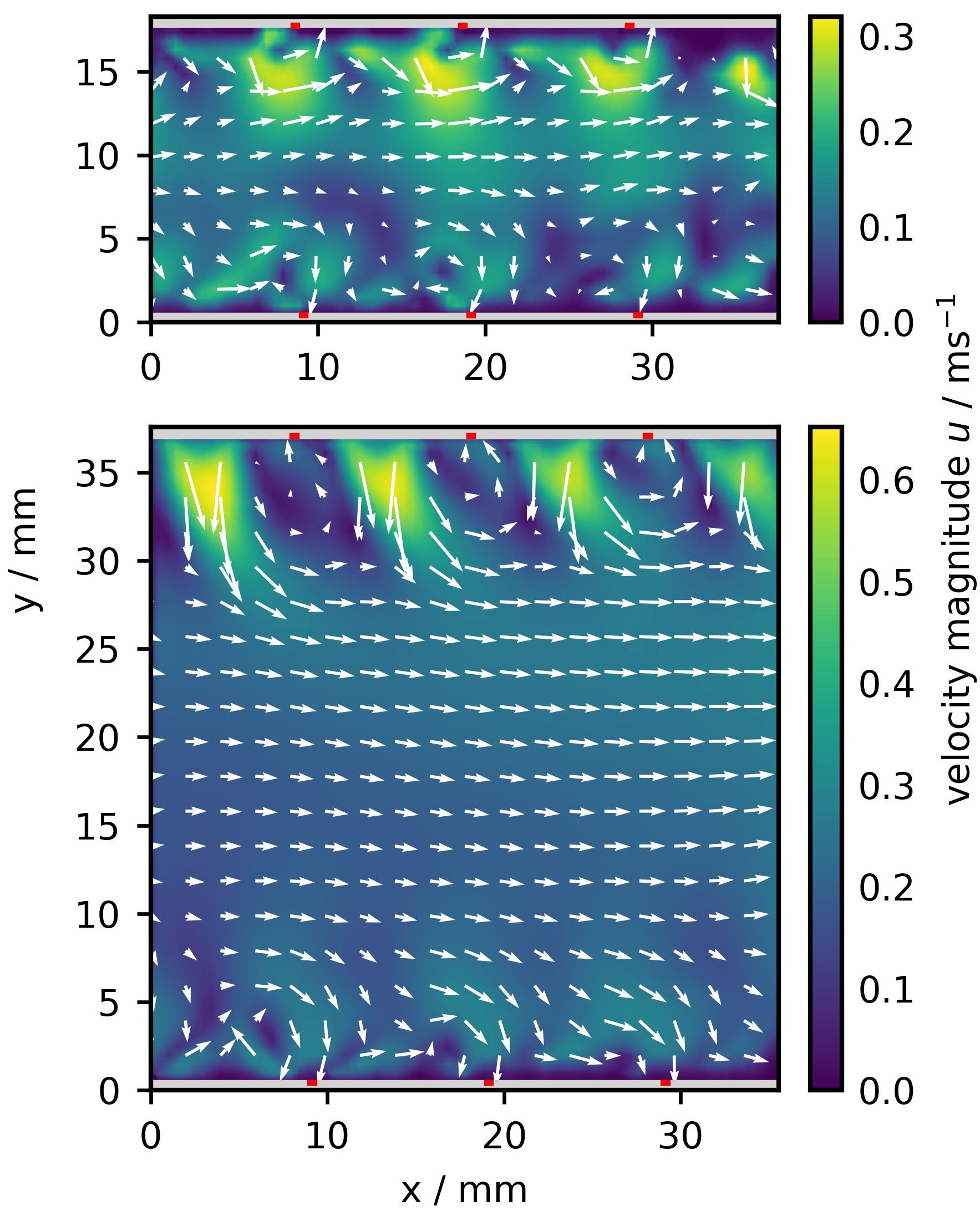}
	\caption{Exemplary flow field results for a gap distance of \SI{16}{mm} (top) and a gap distance \SI{36}{mm} (bottom). The arrow density is artificially reduced by a factor of 4 to enhance the visibility. The electrode is highlighted in light grey colour and the grid positions are marked as red rectangles.  } 
	\label{fig:flowfields}
\end{figure}

Figure \ref{fig:flowfields} presents two exemplary flow fields, with contour colours representing the velocity magnitude for each case individually. The electrode plates are highlighted as horizontal grey rectangles positioned at the top and bottom, while the grid lines are marked with small red rectangles. The horizontal position $x=\,0$ corresponds to the left edge of the observation window. The upper flow field corresponds to a gap distance of \SI{16}{mm} and the lower one has a gap distance of \SI{36}{mm}. In both cases, the external air flow is directed from left to right. Vortices are visible between and above each grid line on both sides. Directly above the grid lines, the flow is drawn towards the surface, while between them, the fluid is pushed away from the surface, resulting in closed vortices. The maximum vertical velocity component $v$ rises with increasing gap distance, reaching approximately \SI{0.3}{ms^{-1}} for a gap distance of \SI{16}{mm} and \SI{0.7}{ms^{-1}} for a distance of \SI{40}{mm}. Compared to the initial velocity in horizontal direction $u_\text{x}$ ($u_\text{y} = \SI{0}{ms^{-1}}$), which is around \SI{0.1}{ms^{-1}}, this represents an increase ranging from \SI{30}{\%} to \SI{68}{\%}. These velocity values enable the transport of $n$-butane towards the surface but should be insufficient for effectively transporting highly reactive species away from the discharge zone due to their short lifetimes \cite{sakiyamaPlasmaChemistryModel2012c}. While the entire flow channel between the two plates is strongly influenced in the \SI{16}{mm} case, a significant part of the flow bypasses the vortex areas in the \SI{36}{mm} case. The ratio between the vortex area height and the gap width is approximately \SI{54}{\%} for \SI{16}{mm} and \SI{36}{\%} for \SI{36}{mm} gap width. In both flow fields, there is no change of the vortices along the flow direction visible, which means that their interaction does not result in spatial consecutive effects. However, the vortices originating from the lower SDBD appear less prominent. These flow fields differ from those observed in our previous study conducted under quiescent air conditions. Here, the vortices appear to be more bent in the direction of flow. We also suggested that displacements of the two opposing grids at one plate, caused by small deviations during manufacturing, could be responsible for the found asymmetric behaviour \cite{boddeckerInteractionsFlowFields2023a}.

\begin{figure}[hbt]
    \centering
    \includegraphics[width=\linewidth]{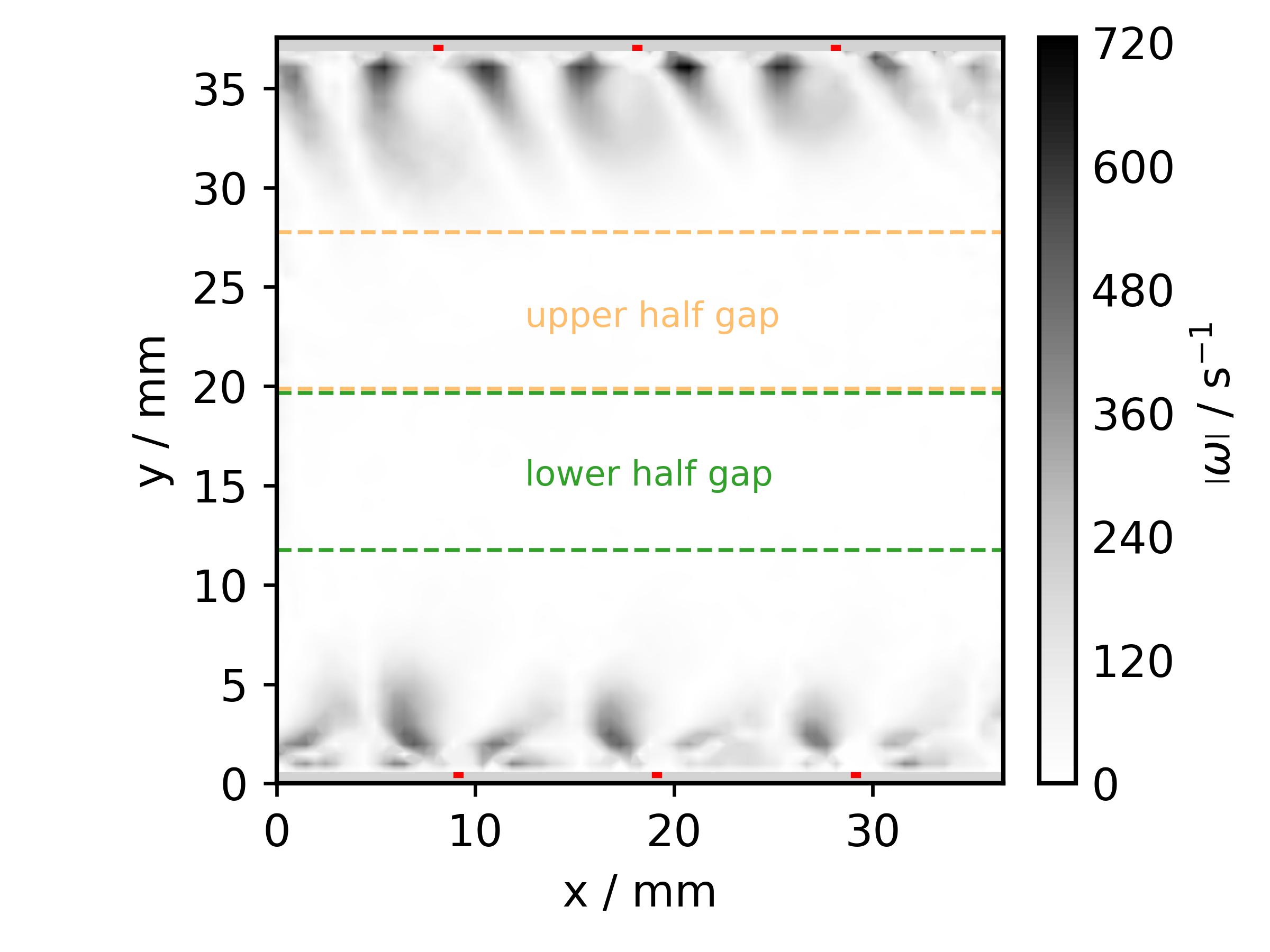}
	\caption{2D absolute value of the vorticity for the gap distance of \SI{36}{mm}. The electrode is highlighted in light grey colour and the grid positions are marked as red rectangles. The borders of integration areas for figure \ref{fig:vorticity_profiles} are marked additionally with horizontal lines.}
	\label{fig:vorticity}
\end{figure}

As discussed above, we use the vorticity to identify vortical structures and to use its absolute values for getting a measure of vortical strength in the flow field. Figure \ref{fig:vorticity} shows the absolute vorticity for the gap distance of \SI{36}{mm}. This depiction highlights the position and strength of the vortices independent of their rotational orientation. Also, here the asymmetry between the top vortical structures and those at the bottom are visible. 

\begin{figure}[hbt]
    \centering
    \includegraphics[width=\linewidth]{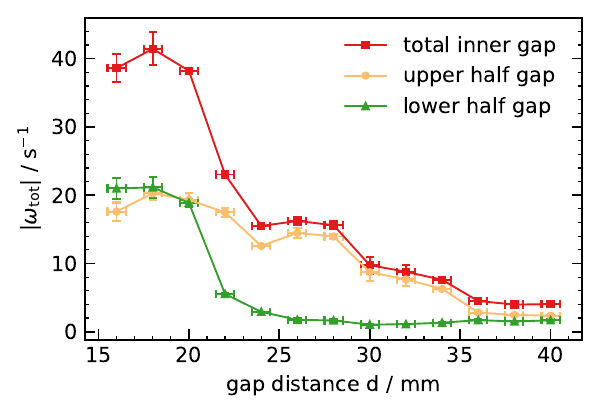}
	\caption{Spatially integrated absolute vorticity $|\omega_{\mathrm{tot}}|$ as a function of the gap distance. The integration window is kept constant to different parts of the area between the electrodes at a minimum gap distance.}
	\label{fig:vorticity_profiles}
\end{figure}

\begin{figure*}[hbt]
    \centering
    \includegraphics[width=\textwidth]{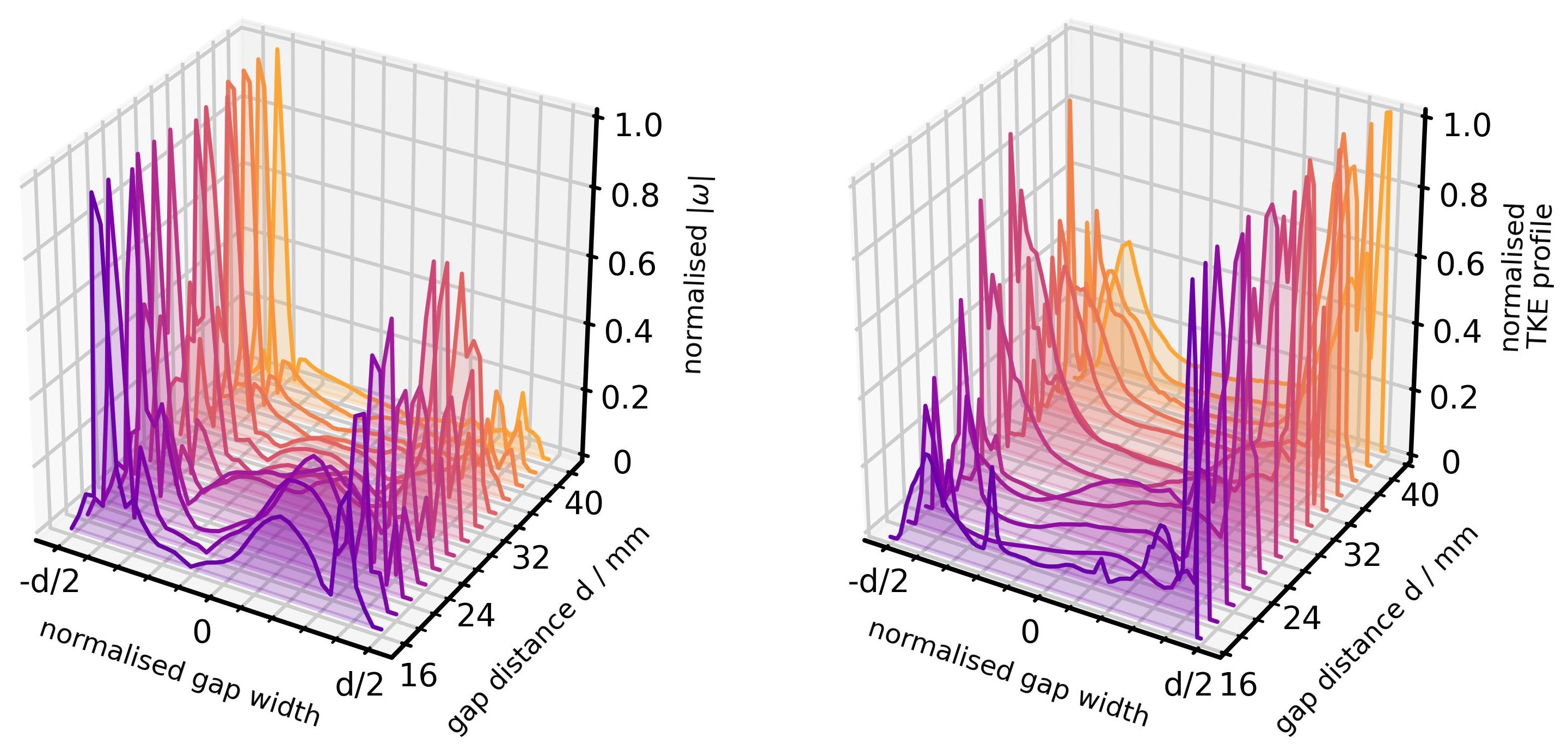}
	\caption{3D graphs for the vorticity (left) and the turbulent kinetic energy (right). Each profile height is normalized to its individual peak value. Additionally, all profile widths (gap widths) are normalized to a uniform gap width from -d/2 to d/2 highlighting relative changes.}
	\label{fig:3dprofiles}
\end{figure*}

In figure \ref{fig:vorticity_profiles} the total absolute vorticity $|\omega_{\mathrm{tot}}|$ across different regions is illustrated. For the analysis, a fixed spatial integration window is utilized. While the total inner gap refers to the area of the flow field at a gap distance of \SI{16}{mm}, the upper half gap and the lower half gap correspond to the halved window with its associated sides. It is important to mention that the window area remains constant for the gap distance variation and is consistently positioned at the centre of the gap. The integration areas are illustrated in figure \ref{fig:vorticity} for a gap distance of \SI{36}{mm}. For \SI{16}{mm} the sum of both displayed areas covers the entire gap between the electrode plates.

The figure shows that a high value region is present for short gap distances, which rapidly decays for gap sizes above \SI{20}{mm}. The vortex size is not increasing proportionally to the gap distance. The strong decay is explained by the fixed analysis window, because the vortices are moving out of this section with growing gap distances. The peak value positions nearly match with the positions of the weak peaks found in the conversion results in figure \ref{fig:conversion_comp}. The decay of $|\omega_{\mathrm{tot}}|$ is more rapid compared to the conversion. One potential reason for the conversion's higher stability could be the increasing gas velocities within the vortices at greater gap distances. While the total vorticity increases with higher applied repetition frequencies and higher resulting gas velocities, the total vorticity rises without reaching the middle gas gap. The slow decay of the conversion cannot be fully quantified yet, and additional factors could play a role. The slight discrepancies in the peak positions by \SI{1}{mm} - \SI{2}{mm} may be attributed to differences between the setups. Notably, while a flow conditioner is utilized for the PIV experiment, the same incoming flow conditions do not apply to both setups. Consequently, the incoming flow in the conversion experiment may exhibit turbulent structures, even without the discharge operation.  Despite these variations, comparing both peak positions allows for a correlation between enhanced relative conversion and increased vorticity at these specific gap distances. This suggests that vorticity serves as a scalable measure of gas mixing in this context.

The investigation of the middle region of the gap seems promising, and therefore the following analysis focuses on this area with higher detail.
Since the single vortices exhibit rough symmetry in the horizontal direction, as visible in figure \ref{fig:vorticity}, a vertical line profile for comparison is feasible. In the following, we selected a horizontal position of approximately \SI{18}{mm} for the line profiles, as it intersects the centre of the vortices on both sides. 

Figure \ref{fig:3dprofiles} shows three-dimensional plots of TKE as well as the vorticity profiles as a function of the normalised gap width and gap distance. Both profiles are normalized to their individual maximum values. Normalising each profile width to its corresponding gap width allows relative structural size comparisons and highlights the flow structures between the gap. The asymmetry between the vortex regions close to the surfaces is also visible here.  The peaks at $-d/2$ are in general more prominent than at $d/2$, which correspond to the top electrode. Their peak width stays mainly constant with respect to the normalized gap distance. The inner structures show a local maximum at \SI{20}{mm} in both plots. This peak can be matched with the ones in the conversion results as well and also supports the idea of enhanced gas mixing at these distances. This result shows that the inner turbulent fluid structures are an important factor for the gas mixing and consequently for the conversion process.

\section{Conclusion and Future Work}
\label{sec:conclusion}

The correlation between induced flow fields from a surface dielectric barrier discharge system and the gas conversion process of $n$-butane is explored. Conversion is measured with flame ionization detectors and the flow fields are calculated by the use of particle image velocimetry. 
Varying the gap distance between two SDBD electrode plates for three different $n$-butane mole fractions revealed local peaks in relative conversion around gap distances of \SI{16}{mm} to \SI{22}{mm}. The relative conversion is decaying gradually with rising distances, but constant energy density. This finding indicates the presence of additional spatially dependent effects. The flow field results show distinct vortex structures at the top and bottom electrodes, evolving in size and form with increasing gap distances. An analysis of the vorticity in the middle gas gap region corresponds with the identified peaks in conversion. The faster decay of the vorticity can be partially attributed to the fixed integration window in the middle, while the total vorticity increased with greater gap distances due to the repetition frequency rise. The rapid decay cannot be fully quantified at this stage. Vorticity and turbulent kinetic energy analyses provided insights into these structures' characteristics and their impact on gas mixing. The vorticity field identifies the number and strength of the vortices clearly, which have symmetrical structures in the horizontal direction and do not show consecutive processes. Line profiles through the centre of the vortices showed peaks in the middle gap region, correlating with the observed peaks in conversion. This congruence shows the correlation between SDBD gas conversion processes and its induced flow fields due to gas mixing enhancement. 

Future investigations will be necessary to determine the effect of polarity alteration between the electrodes to assess the presence of an electric field between the electrodes. This will potentially reveal the effect of asymmetries in streamer dynamics to the flow field.  
Complementary simulations will support uncovering the complex interaction of the induced flow fields. 

\section*{Acknowledgements}
The investigation presented in this paper received financial support from the German Research Foundation (DFG) through projects A7 and A5 within the Collaborative Research Centre SFB 1316 (project number 327886311), ``Transient atmospheric pressure plasmas - from plasmas to liquids to solids."

\section*{References}
\providecommand{\newblock}{}

\end{document}